\documentclass[apl,onecolumn,showpacs,groupedaddress,nofootinbib,notitlepage]{revtex4-1}

\usepackage{graphicx}
\usepackage{url}
\usepackage{natbib}
\usepackage{amsmath,amssymb,amsthm,bm,mathtools,amsfonts,mathrsfs,bbm} 
\usepackage{siunitx}

\usepackage{bbold}
\usepackage{dsfont}
\usepackage{latexsym}
\usepackage{fancyhdr}
\usepackage{array}

\usepackage[babel]{microtype}  
\usepackage{centernot}

\usepackage{gensymb}
\usepackage{dsfont}
\usepackage{tabularx}

\DeclareMathOperator{\Tr}{Tr}

\newcommand{\bra}[1]{\left\langle #1 \right|}
\newcommand{\ket}[1]{\left| #1 \right\rangle}

\newcommand{\ketbra}[2]{\left|#1\middle\rangle\middle\langle#2\right|}

\def\be{\begin{eqnarray}}
\def\ee{\end{eqnarray}}

\usepackage{changes}
\usepackage{comment}

\usepackage{etoolbox}
\patchcmd{\section}
  {\centering}
  {\raggedright}
  {}
  {}

\begin{document}

\title{Heralded Amplification of Path Entangled Quantum States}

\author{F.~Monteiro$^1$, E.~Verbanis$^1$, V.~Caprara Vivoli$^2$, A.~Martin$^1$, N. Gisin$^1$, H.~Zbinden$^1$ and R.~T.~Thew$^1$}


\affiliation{$^1$Group of Applied Physics, University of Geneva, Switzerland}
\affiliation{$^2$ QuTech, Delft University of Technology, Lorentzweg 1, 2611 CJ Delft, The Netherlands }

\begin{abstract}
Device-independent quantum key distribution (DI-QKD) represents one of the most fascinating challenges in quantum communication, exploiting concepts of fundamental physics, namely Bell tests of nonlocality, to ensure the security of a communication link. This requires the loophole-free violation of a Bell inequality, which is intrinsically difficult due to losses in fibre optic transmission channels. Heralded photon amplification is a teleportation-based protocol that has been proposed as a means to overcome  transmission loss for DI-QKD. Here we demonstrate heralded photon amplification for path entangled states and characterise the entanglement before and after loss by exploiting a recently developed displacement-based detection scheme. We demonstrate that by exploiting heralded photon amplification we are able to reliably maintain high fidelity entangled states over loss-equivalent distances of more than 50~km.

\end{abstract}

\maketitle

\section{Introduction}

The distribution of entanglement is a key resource for quantum communication~\cite{Gisin2007}. Single-photon entanglement, also called path entanglement, represents possibly the simplest form of entanglement to generate~\cite{Tan1991,VanEnk2005}, and one that lies at the heart of some of the most efficient quantum repeater architectures~\cite{Sangouard2011}. Nonetheless, in the context of quantum communication, a means of measuring these states in distributed scenarios has proven difficult. The measurement, and subsequently the characterisation, of these states has been recently addressed by adopting the relatively old idea of what we call displacement-based detection~\cite{Banaszek1999,Kuzmich2000,Bjork2001,Bjork2004}. This scheme combines aspects of discrete- (photon counting) and continuous- (local oscillator) variable detection and has recently been used to perform entanglement witness~\cite{Monteiro2015} and detection loophole-free EPR steering~\cite{Guerreiro2016} experiments.

Transmission loss is critical for all quantum communication scenarios, however, heralded photon amplification (HPA)~\cite{Ralph2009}, or noiseless linear amplification, provides a potential solution to mitigate its impact. This is a teleportation-based protocol that has been experimentally studied in the case of both polarisation~\cite{Kocsis2012} and time-bin~\cite{Bruno2016} qubits as well as for single photons~\cite{Xiang2010,Osorio2012} and in the continuous variables regime~\cite{Grangier2010,Fuwa2014}. In the context of entangled systems, loss represents a fundamental limit for the detection loophole-free Bell test~\cite{Hensen2015,Giustina2015,Shalm2015} needed for device-independent quantum key distribution (DI-QKD)~\cite{Acin2006}. Heralded amplification has been proposed as a means to overcome loss in the critical case of DI-QKD~\cite{Gisin2010,Pitkanen2011,Meyer-Scott2013}. More generally, it can also be seen as an entanglement distillation protocol, whereby the herald for the amplifier announces, or selects, a subset of states that have a higher degree of entanglement than before. First experiments in this direction for path entanglement relied on joint measurements~\cite{Xiang2010,Osorio2012} due to the difficulty in measuring these states in a distributed scenario, i.e. with local measurements. Continuous variable entangled systems, using homodyne measurements, have also used HPAs to distil entanglement, demonstrating the flexibility of this concept~\cite{Chrzanowski2014,Lvovsky2015}. 

Here, we bring together all three of these concepts: heralded path entanglement; displacement-based detection, and heralded photon amplification, all operating at telecom wavelengths. We demonstrate how heralded photon amplification can be exploited to overcome transmission loss, up to a loss-equivalent distance of around 50~km. The initial and final states are measured using a displacement-based detection scheme and characterised by their Fidelity with respect to a maximally entangled state.

\section{Concept and Theory}

\begin{figure}[!t]
\centering
\includegraphics[scale=0.35]{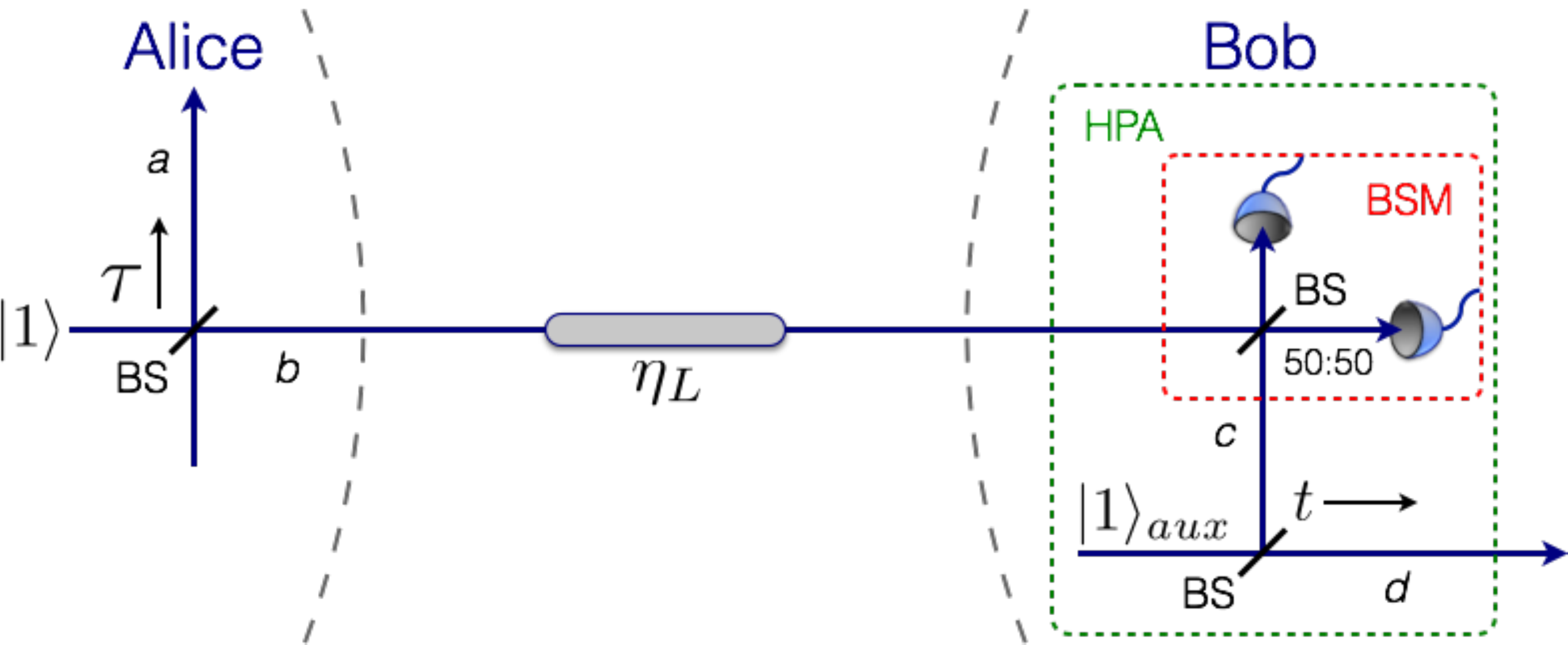}
\caption{Concept combining heralded path entanglement and heralded single photon amplification: Alice prepares an initial path entangled state between the modes $a$ and $b$, which she sends to Bob through a lossy channel $\eta_L$. Bob's heralded photon amplifier (HPA) works by impinging an auxiliary photon on a highly-transmissive ($t$) beamsplitter (BS) before performing a Bell state measurement (BSM) between his half of the entangled state ($b$) and part of the auxiliary photon ($c$). A click in one of the BSM detectors heralds the loss-reduced state into modes $a$ and $d$. }
\label{setup_concept}
\end{figure}

Fig.~\ref{setup_concept} illustrates the scheme for generating and distributing path entanglement as well as how we incorporate the heralded photon amplifier. One should note the conceptual similarity to an entanglement swapping scheme. In the following, we elaborate on the theoretical description of this system and how we can characterise and compare the entanglement before and after the HPA. Alice starts by sending a photon through a beamsplitter (BS) with splitting ratio $\tau:1-\tau$, and creates the entangled state $\ket{\psi(\tau)} =  \sqrt{\tau}\ket{10}_{ab} + \sqrt{1-\tau}\ket{01}_{ab}$ in the photon number basis. Part of the entangled state is sent to Bob via a lossy link with transmission  $ \eta_{L} $. The initial state, after passing through this link, can be written as:
\begin{equation}
\rho_{i} = (1-\eta_{L})(1-\tau)\ketbra{00}{00}+ (\tau + \eta_{L}(1-\tau))\ketbra{\psi_{i}}{\psi_{i}}
\end{equation}
where $\ket{\psi_{i}}$ is given by
\begin{equation}
\ket{\psi_{i}} = \frac{\sqrt{\tau}\ket{10}_{ab} +  \sqrt{\eta_{L}(1-\tau)}\ket{01}_{ab}}{\sqrt{\tau + \eta_{L}(1-\tau)}},
\end{equation}
describing the state shared between modes $a$ and $b$ as indicated in Fig~\ref{setup_concept}.

Bob then uses a HPA~\cite{Ralph2009} to compensate the lossy link on his part of the entangled state. The heralded photon amplifier works in the following way: Bob uses an auxiliary photon that impinges on a BS that has a transmission $t$ in the direction of Bob's measuring device and a reflection $1-t$ towards the Bell state measurement (BSM). Both the path entangled state and the reflected auxiliary photon interfere on the BS, whereafter, the detection of a single photon heralds a successful BSM, and hence, amplification process, such that Alice and Bob can now recover the entanglement in their shared state. One can see this as entanglement swapping, where Alice prepares an entangled state in modes $a$ and $b$ while Bob prepares one in modes $c$ and $d$. The difference here is that just a single detection for the BSM on modes $b$ and $c$, results in an entangled state shared between modes $a$ and $d$.

A successful BSM heralds a final state of the form:
\begin{equation}
\rho_{f} = \frac{1}{N}\{\lambda\ketbra{00}{00} + (N-\lambda)\ketbra{\psi_f}{\psi_f}\}
\label{final_state_model_simple}
\end{equation}
where the entangled component of the state shared by Alice (mode $a$) and Bob (mode $d$) is given by:
\begin{equation}
\ket{\psi_f} = \frac{\sqrt{\tau(1-t)}\ket{10}_{ad} +  \sqrt{t\eta_{L}(1-\tau)}\ket{01}_{ad}}{\sqrt{\tau(1-t)+t\eta_{L}(1-\tau)}}.
\label{pij_behaviour}
\end{equation} 
Note that using the other outcome (detector) of the BSM produces a phase-flipped version of this state. $\lambda = (1-\tau)(1-t) $ defines the loss, or vacuum, component of the heralded state and $N = (1-\tau)\eta_{L} t + (1-t)$ gives the state normalisation. From this we can determine the probability of a successful BSM, $P_{BSM}=N/2$, for either of the BSM detectors. We note that $\rho_{f}$ holds for binary detectors; the use of photon-number-resolving detectors can further improve the performance of the HPAs~\cite{Osorio2012,Bruno2016}. The HPA acts on the initial lossy state in two ways. We see from Eq.~\ref{final_state_model_simple} where, by varying the HPA's transmission parameter $t$, we can make $\lambda$ small, thus overcoming the loss. If we look at Eq.~\ref{pij_behaviour}, we see that loss can also affect the degree of entanglement, which again, can be overcome by varying $t$ and $\tau$.

We characterise the initial ($i$) and final ($f$) states $\rho_{i,f}$ by employing a measurement scheme~\cite{Monteiro2015,Vivoli2015,Guerreiro2016} that uses small displacements $\cal{D}$$(\alpha)=e^{\alpha a^\dag -\alpha^\star a}$ operating on each of Alice's and Bob's modes~\cite{Paris1996}, followed by binary detectors. The resulting detection click ($C$) and no-click ($0$) probabilities are given by: 
 \begin{eqnarray}
P_0 =  \Tr{[\rho \ketbra{\alpha}{\alpha} ]} \hspace{2cm}    P_C = 1 - P_0,
 \label{eq:pclick} 
 \end{eqnarray}
where $\rho$ represents the reduced state for either Alice or Bob.  We can use these measurements to obtain the conditional probabilities $P_{00}, P_{C0}, P_{0C}, P_{CC}$, which are dependent on the measurement (phase and amplitude) settings of both Alice's and Bob's displacements, which gives us access to approximate $\sigma_z$, $\sigma_x$ and $\sigma_y$ Pauli matrix operators~\cite{Monteiro2015,Vivoli2015,Guerreiro2016}. We also note that all measurement outcomes are considered, i.e. there is no post-selection, nor do we truncate the Hilbert space as the $P_C$ terms include contributions from higher photon-numbers. From these probabilities we can evaluate quantities such as the Fidelity which take the standard form $F_k = \bra{\psi}\rho_{k}\ket{\psi}$, where $\ket{\psi}$ is the maximally path entangled state and $k$ denotes the initial (lossy) or final (amplififed) state.

The Fidelity is obtained by considering the joint probabilities $P_{mn}$ to have outcomes $m$ for Alice and $n$ for Bob and is given by 
\begin{equation}
F = \vert d\vert _{\alpha=0.7} + 0.5\left[P_{0C} + P_{C0}\right]|_{\alpha=0},
\label{F}
\end{equation}
where the off-diagonal coherence term $d=\bra{01}\rho\ket{10}$ is a function of all $P_{mn}$ when the displacements $\cal{D}$$(\alpha_{A})\otimes$$\cal{D}$$(\alpha_{B})$ are applied ($|\alpha| \approx0.7$ providing the optimal displacement).

The probabilistic nature of the BSM for the amplification imposes a trade-off, as high values of $F_f$ implies low heralding rates. It is possible to obtain $F_f \rightarrow 1$ when $\tau = \frac{\eta_L  t}{1+\eta_L (1-t)}$ and $t\rightarrow 1$, however, the probability of success also then tends to zero. Note that in the case of $\tau=1$ or $t=1$ the state is separable. Hence we need to verify both that the protocol increases the Fidelity and that it preserves the entanglement. This is done by comparing the measured Fidelities $F_f$ to the maximum Fidelity that can be obtained with a separable state, $F_{sep}$, i.e. where $\rho$ stays positive under partial transposition in the \{0, 1\} subspace \cite{PPT1,PPT2}.

The separable Fidelity $F_{sep}$ also takes into account the probability of having two photons at Alice $P_{2a}$ and at Bob $P_{2b}$ (or $P_{2d}$ in the case where Bob applies the HPA): 
\begin{equation}
F_{sep} = \left[\frac{P_{C0} + P_{2a}}{2} + \frac{P_{0C} + P_{2b}}{2} + \sqrt{P_{00}(P_{CC} + P_{2a} + P_{2b})}\right] |_{\alpha=0}
\label{F_Separable}
\end{equation}
The probabilities $P_{2a}$ and $P_{2b}$ ($P_{2d}$) are determined from the $g^2(0)$ measurements performed by Alice and Bob.

\section{Experiment}

\begin{figure*}[!t]
\centering
\includegraphics[scale=0.38]{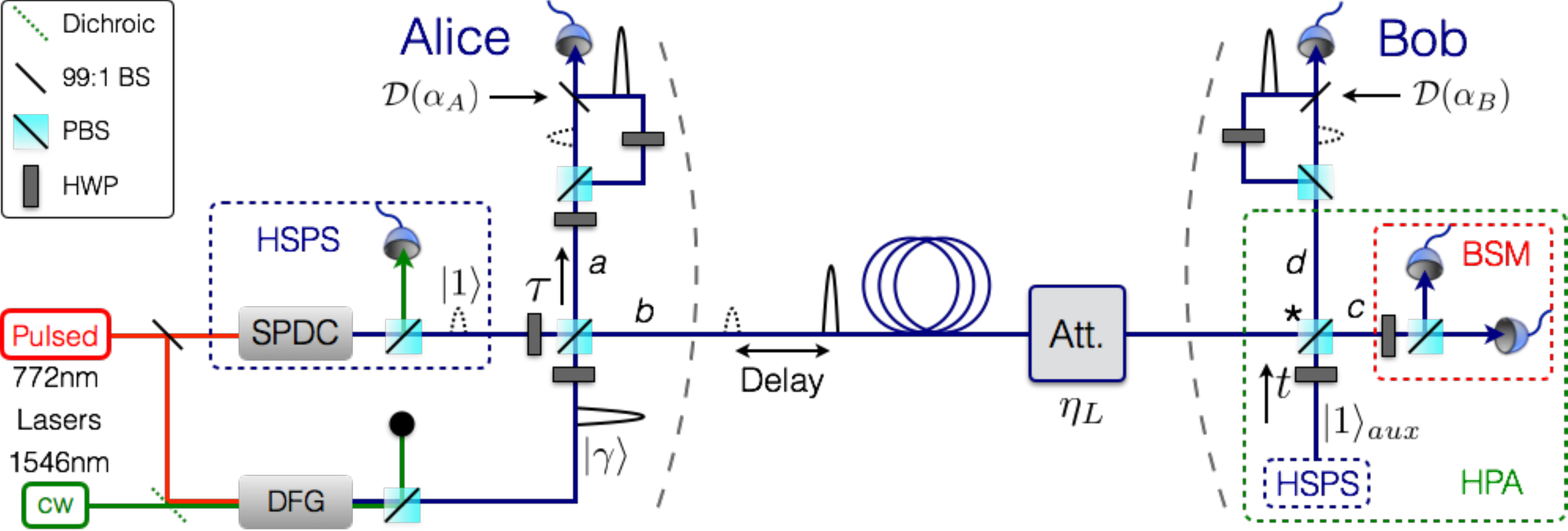}
\caption{Experimental schematic: A heralded single-photon source (HSPS) is used to herald entanglement between modes $a$ and $b$, which is coupled into fibre and sent to Bob. Transmission loss is simulated by attenuating (Att.) mode $b$. A second HSPS provides an auxiliary photon for the heralded photon amplifier (HPA). A detection at the Bell state measurement (BSM) heralds the final state between Alice (mode $a$) and Bob (mode $d$). Difference-frequency generation (DFG) is used to generate a weak coherent state to perform the displacement operations at the 99:1 beamsplitter (BS) before detection with single photon counters.  Legend indicates dichroic mirror (DM), polarisation beamsplitter (PBS) and half-wave plates (HWP). See text for detail.}
\label{setup_ampli_2}
\end{figure*}

A simplified experimental schematic is shown in Fig.~\ref{setup_ampli_2}. Two heralded single photon sources (HSPS) are used to create both the path entangled states on Alice's side, between modes $a$ and $b$, and the auxiliary photons on Bob's side, which are also non-maximally entangled between modes $c$ and $d$. The HSPSs are realised by detecting one photon from each pair generated in a double pass spontaneous parametric down-conversion (SPDC) scheme, although to simplify Fig.~\ref{setup_ampli_2}, these are shown as independent sources. The sources are based on a PPKTP nonlinear crystal pumped with a 76\,MHz pulsed (ps) laser at 772 nm that generates pure photons ($>$~90\%) in telecom-band without spectral filtering~\cite{Bruno2014a}. The degree of entanglement for Alice's initial heralded entangled state is determined by rotating a half-wave plate (HWP) before a polarisation beamsplitter (PBS), corresponding to a BS ratio $\tau:1-\tau$. A gated InGaAs single photon detector, with a detection efficiency of $\sim 25\%$, heralds 50~kHz of entangled states, with one half directed locally to Alice's measurement setup and the other coupled into fibre and sent to Bob. The transmission $\eta_L$ through this link is attenuated (Att.) to simulate distance.

On Bob's side we use a modified set-up for the HPA compared to Fig.~\ref{setup_concept}. The second HSPS generates the auxiliary photon which passes through another HWP and PBS, denoted with $*$, which determines the BS splitting ratio $t:1-t$ for the HPA. As explained previously, $\tau$ and $t$ can be used to optimise the Fidelity of the final amplified state as a function of channel transmission. Bob's half of the initial entangled state is sent straight through PBS($*$) and combined with the reflected, and orthogonally polarised, component ($1-t$) of the auxiliary photon. The BSM then consists in rotating these, via another HWP to erase the polarisation information, and interfering them at another PBS. A single detection - a successful BSM - then heralds a loss-reduced version of the state that was shared between modes $a$ and $b$, into Alice and Bob's output modes $a$ and $d$.

Path entangled state analysis is performed via a displacement-based measurement scheme \cite{Bjork2004,Monteiro2015,Guerreiro2016}. Experimentally, this is realised by interfering the coherent state $\ket{\gamma}$ with each mode of the entangled state and then using single photon counting detectors. This approach can be seen as somewhere in between continuous and discrete variable measurement schemes, where the weak  coherent state, $|\alpha| \approx 0.7$, gives us access to phase dependent measurements in the photon counting regime, i.e. to make measurements in the $|0\rangle \pm |1\rangle$ basis. An important requirement for this scheme is that there are no phase fluctuations or drifts between the coherent state and the path entangled state to be analysed,  e.g. from path-length dilation of transmission fibres due to changes in temperature. To ensure this, the coherent state is input to the system at the first PBS on Alice's side so that it can co-propagate with the entangled state; the two states are orthogonally polarised. The use of polarisation allows us to reflect the coherent state towards Bob's output mode $d$ at the PBS($*$), thus co-propagating with the heralded, and amplified, state in mode $d$. 

As illustrated in Fig.~\ref{setup_ampli_2} the coherent state used for the displacement-based measurements is generated using difference-frequency generation (DFG) between the pulsed pump laser at 772 nm and a cw laser at the same wavelength (1546\,nm) as the heralding photon, using a PPLN nonlinear crystal~\cite{Bruno2014b}. Another HWP and PBS couples this into modes $a$ and à$b$, with orthogonal polarisation to the entangled state and with a fixed temporal delay. This delay allows us to use gated detectors for the BSM that only operate when the entangled state is expected and not the coherent state, thus avoiding false detections that would introduce errors in the final state. Finally, the two states are sent to unbalanced Mach-Zehnder interferometers, where a PBS directs them into the two different paths to erase the timing information. In the interferometers the polarisation of the coherent state is rotated, by a HWP, before interfering with the entangled state at the 99:1 fibre BSs, which correspond to the displacement operations. The interferometers are actively, and independently, stabilised and piezo-actuated mirrors allow us to perform measurements with well-defined phase differences between Alice and Bob. 

A delay line between the two HSPSs is used to ensure that the photons arrive at the BSM at the same time and another delay between these and the DFG ensures their synchronous arrival at the 99:1 BSs for the displacement operations. 
The indistinguishability, or overlap, between the photons and the coherent state is essential for the displacement operation and is measured directly from several $P_C$ measurements: without the coherent state; without the reduced (entangled) state of Alice and Bob, and then with both.  This provides a more direct, and faster measurement than a HOM interference measurement; we determine an overlap of $0.90 \pm 0.02$.

We have an initial entanglement heralding rate of around \SI{50}{kHz}, however, the final, amplified, heralding rates are significantly lower due to the probabilistic nature of the two SPDC sources, which simultaneously generate $\approx 30$ path entangled states and auxiliary photons per second. In the case with amplification, the heralding rate is also dependent on the loss, achieving rates of \SI{0.25}{Hz}, \SI{0.19}{Hz} and \SI{0.15}{Hz}, respectively, for decreasing $\eta_{L}$. The transmission $\tau=0.6$ was used to create the initial path entangled state to be amplified, while the transmission of the auxiliary photon through the PBS($*$) was set to $t=0.93$. These two values were maintained for all measurements, as they provide a reasonable trade-off between measurement time and Fidelity $F_f$ for all tested cases.

\section{Results}

\begin{figure*}[!t]
\centering
\includegraphics[scale=0.35]{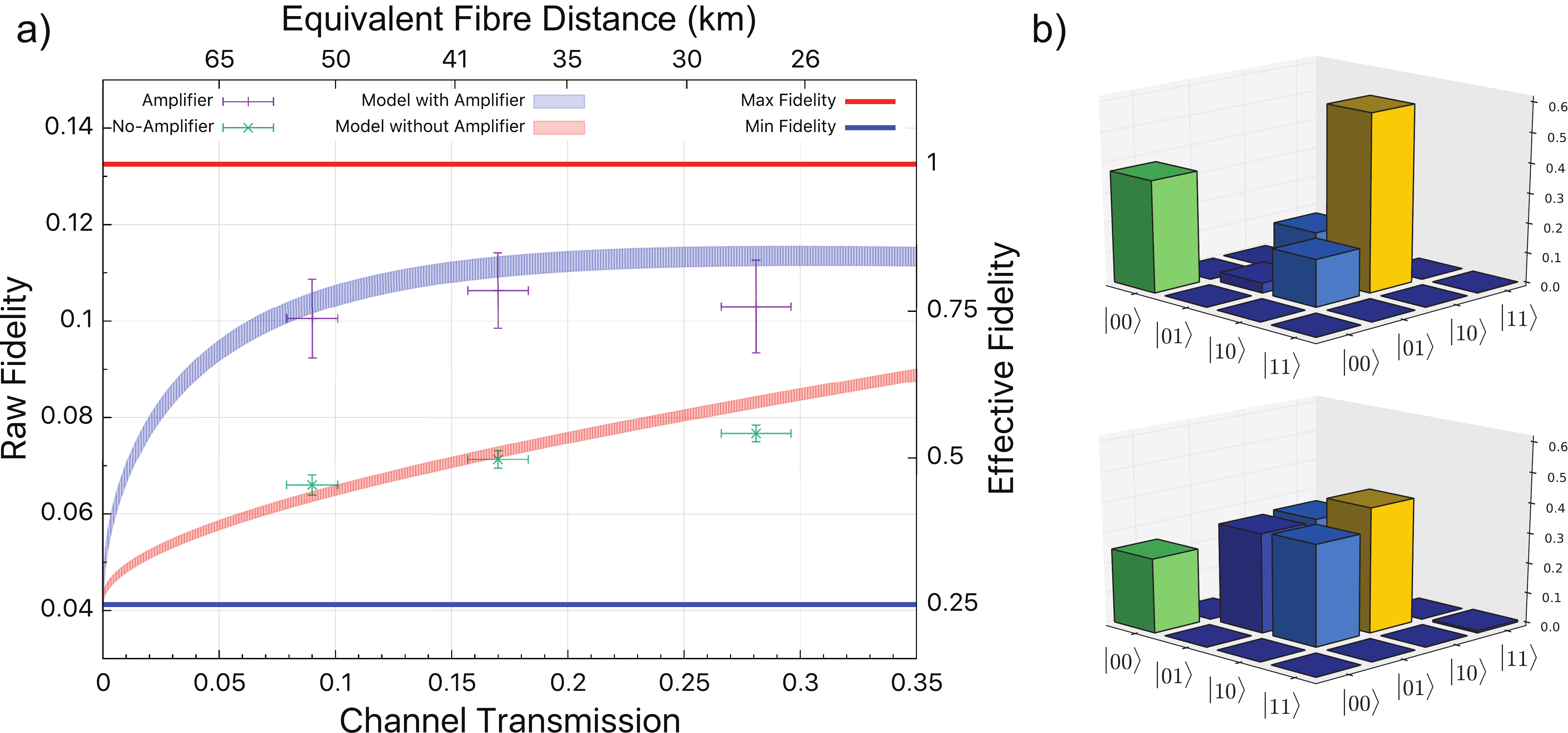}
\caption{a) Measured Fidelities with and without the amplifier for different channel transmissions. The upper (blue) and lower (pink) bands are calculated values of the Fidelity, with the maximally entangled state, using the expected state before and after amplification. The top axis shows the fibre distance that has equivalent transmission of our channel. The right axis shows the expected Fidelities we would obtain in the case Alice and Bob use single photon detectors with unit efficiency. b) Reconstructed density matrices for the states before (top) and after (bottom) the HPA at  $\eta_L = 0.09$.}
\label{Fidelity_measured}
\end{figure*}

We characterise and compare the Fidelities for the entangled state, with and without the HPA, for total transmissions $\eta_{L}$ of approximately 0.28, 0.17 and 0.09, corresponding to equivalent transmission distances in standard telecom fibre (0.2 dB/km) of around 30, 40 and 50\,km. The main results are shown in Fig.~\ref{Fidelity_measured}a. In the current configuration we use InGaAs detectors, with detection efficiencies around 25\,\%, and the corresponding raw Fidelity is shown on the left vertical axis. On the right vertical axis we re-scale these results, based on unit detection efficiency, to help differentiate the effect of detection efficiency from the performance of the state generation, displacement-based measurements, and HPA.

The Fidelities are obtained from the measured joint probabilities $P_{mn}$, following Eq.~\ref{F} and Eq.~\ref{F_Separable}, for displacement amplitudes $\vert\alpha_{a,b}\vert=0$ and $\vert\alpha_{a,b}\vert \approx 0.7$ and the probabilities of having more than one photon in each path, due to double pairs of the SPDC sources. As an example, Table~\ref{table_Pmn} shows the raw measured values at $\eta_L = 0.09$, while $P_{2a}, P_{2b} < 10^{-4}$. Using the measurement results for $P_{mn}$ when $\alpha \approx 0.7$, we can calculate the off-diagonal coherence term $d = (4.8 \pm 1.6 )\times 10^{-2}$. The error bars for the experimental results in Fig.~\ref{Fidelity_measured}a for the amplified case are quite large and are dominated by the measurement of the off-diagonal coherence term $d$, which is a function of all the probabilities $P_{mn}$ and terms exponential in $|\alpha|^2$; $|\alpha |$ can vary over time resulting in errors larger than just the statistical case. We limit these measurements to around $2.6\times 10^{3}$ events, compared to  $38\times 10^{3}$ events without the coherent state, for each attenuation. Measurements with the coherent state, while having a higher $P_C$, are less stable as the polarisation of the coherent state is not actively stabilised before injecting into the set-up and can rotate, as such this is re-aligned and we take many measurements of around one hour and accumulate these over a few days. The measurements without coherent state do not suffer from any stability problem and hence these are simply left as well for several days, giving rise to better statistics. Fig.~\ref{Fidelity_measured}a also shows the theoretically expected Fidelities with and without amplification as a blue (upper) and pink (lower) curves, respectively. They take into account the coupling loss of the photons, the detection and transmission efficiencies for Alice and Bob as well as any imperfections in the spectral overlap between the heralded photons and the coherent state. The error bars of the model are determined by considering the uncertainties on the experimental parameters, such as the measured BS ratios, and values for the transmission and detection efficiencies. 
\begin{table}[!h]
\centering
\setlength{\tabcolsep}{12pt}
\renewcommand{\arraystretch}{1.5}
\begin{tabular}{cc||c|c|c|c}
 & & $P_{00}$ & $P_{0C}$ & $P_{C0}$ &  $P_{CC}$ \\ \hline \hline
& $\vert\alpha_{a,b}\vert=0$           & $0.9070 \pm 0.0001$  & $ (4.3 \pm 0.03) \times10^{-3}  $  &  $ (8.8 \pm 0.01) \times10^{-2}$  &  $(4.0 \pm 0.3)\times10^{-5}$	\\ \hline
&  \hspace{3mm}$\vert\alpha_{a,b}\vert \approx 0.7$         & $0.358 \pm 0.001$    & $0.219 \pm 0.001$                       &  $0.251\pm 0.001$                      & $0.171 \pm 0.001$ \\ \hline \hline
HPA & $\vert\alpha_{a,b}\vert=0$   & $0.894\pm0.002$      & $0.047\pm0.001$                         &  $0.059\pm0.001$                       & $(8.8 \pm 1.5)\times10^{-4}$ \\ \hline
HPA &  \hspace{3mm}$\vert\alpha_{a,b}\vert \approx 0.7$ & $0.374\pm0.010$      & $0.219\pm0.008$                         &  $0.236\pm0.008$                       & $0.170 \pm 0.007$ \\ 
\end{tabular}
\caption{Measured joint probabilities $P_{mn}$ without and with the HPA at $\eta_L = 0.09$.}
\label{table_Pmn}
\end{table}

One can observe that the measured Fidelities not only match the theory in both cases, but a clear increase in Fidelity is achieved after amplification. While a significant increase in Fidelity is demonstrated, one must also ensure that the final state preserves entanglement. This is realised by comparing the measured Fidelities $F_f$ to the separable state $F_{sep}$, following Eq.~\ref{F_Separable}. We note that $F_{sep}$, which corresponds to the case where the coherence is lost, i.e. we have a mixed state of vacuum, $|10\rangle\langle10|$ and $|01\rangle\langle01|$ terms, is larger than the minimal Fidelity shown in Fig.~\ref{Fidelity_measured}a. The minimal Fidelity corresponds to the case where Bob's photon is completely lost and we approach a mixed state of vacuum and $|10\rangle\langle10|$. The right hand side of Fig.~\ref{Fidelity_measured}b shows the reconstructed density matrices for the initial (top) and final (bottom) states,  assuming unit detection efficiencies, where we more clearly see the effect of the HPA. These results are for a transmission of $\eta_L = 0.09$, which is equivalent to more than 50\,km of fibre transmission. In the case of the initial state, before the HPA, the $|01\rangle_{ab}$ component is greatly attenuated and there is a significant vacuum contribution. After the HPA we see this vacuum contribution is noticeably reduced and we recover a state close to the maximally entangled state. This final state also reveals that the $\tau$ and $t$ parameters could be further optimised to improve the Fidelity.

Table~\ref{table_fsep} compares $F_{sep}$ with the amplified Fidelities  for each measurement with different $\eta_L$. Even though $F > F_{sep}$ for each measurement, the respective confidence levels are only about $1.2$, $2.4$ and $2.2$ standard deviations, respectively.  Nonetheless, by combining all results, it is possible to calculate a probability of around $5\times 10^{-6}$ that $F < F_{sep}$, showing with a high confidence level that the HPA is preserving entanglement. 
\begin{table}[!h]
\centering
\setlength{\tabcolsep}{12pt}
\renewcommand{\arraystretch}{1.5}
\begin{tabular}{c||c|c|c}
$\eta_{L}$        		&  $\approx 0.28$  	& $\approx 0.17$ 	&  $\approx 0.09$ 	\\ \hline
$(F_f - F_{sep})/F_{sep} $  & $0.121 \pm 0.104$  & $0.221 \pm 0.091$ 	&  $0.254 \pm 0.112$ \\
\end{tabular}
\caption{Calculated $(F_f - F_{sep})/F_{sep}$ for each amplified state. The probability that $F<F_{sep}$ for all the combined measurements is $\approx 5\times 10^{-6}$, showing with a high confidence level that the heralded photon amplifier is preserving entanglement.}
\label{table_fsep}
\end{table}

In this demonstration we use standard gated InGaAs SPADs for photon detection, which are limited to around 25\,\% efficiency. While this allows us to clearly demonstrate the potential of this approach, it is not sufficient to overcome the detection efficiency threshold necessary for device-independent protocols. One significant improvement would be to use superconducting nanowire single photon detectors (SNSPD), however, in this configuration we cannot exploit their full potential, in terms of detection efficiency, due to detecting photons from the weak coherent states when no entanglement has been heralded. These detections can blind the detectors, or reduce their effective efficiency, while they recover. Previously, some of us performed a detection loophole-free  EPR steering experiment - a one-sided semi-device-independent protocol - using SNSPDs~\cite{Guerreiro2016}, however the clock rate of the system had to be significantly reduced to match the relatively slow recovery time of the SNSPDs. To exploit the high detection efficiencies of SNSPDs their recovery time has to be reduced \cite{Berggren2016} or a means of operating them in a gated mode needs to be developed \cite{Majedi2012}, which are both ongoing areas of research for this technology.

\section{Conclusion}

We have brought together three concepts: heralded path entanglement generation; displacement-based detection, and heralded photon amplification, demonstrating that high Fidelity entanglement can be distributed over distances equivalent to up to 50\,km. The original proposal for Device-Independent quantum key distribution (DI-QKD), based on heralded photon amplifiers~\cite{Gisin2010}, was for polarisation entangled photons, conceptually, however, path entanglement provides a much simpler implementation. This is the first experimental demonstration of a system suitable for implementing this DI-QKD protocol. The current distance limitation is governed primarily by the probabilistic nature of the SPDC sources and was already alluded to in the initial proposal. Nonetheless, \SI{50}{km} would already be a significant distance for DI-QKD. In combination with recent experiments demonstrating heralded path entanglement and displacement-based detection~\cite{Monteiro2015,Guerreiro2016}, the combination with heralded photon amplification represents a fascinating and flexible toolbox for testing concepts of device-independent protocols and entanglement distribution in quantum networks.

\begin{acknowledgments}
The authors would like to thank N. Bruno and N. Sangouard for useful discussions. This work was supported by the Swiss national science foundation (Grant No. 200021\_159592) and by Nederlandse Organisatie voor Wetenschappelijk Onderzoek (NWO) (VIDI).
\end{acknowledgments}

\bibliography{bib_ent_amp}

\end{document}